\documentclass{emulateapj}
\usepackage{apjfonts}
\usepackage{graphicx}
\usepackage{xspace}

\newcommand{\Msun}      {\mbox{$\rm\,M_{\mathord\odot}$}}

\begin{document}

\lefthead{A Delayed Transition for 4U~1630--47}
\righthead{Tomsick et al.}

\submitted{Accepted by ApJ}

\def\lsim{\mathrel{\lower .85ex\hbox{\rlap{$\sim$}\raise
.95ex\hbox{$<$} }}}
\def\gsim{\mathrel{\lower .80ex\hbox{\rlap{$\sim$}\raise
.90ex\hbox{$>$} }}}

\title{A Delayed Transition to the Hard State for 4U~1630--47 at the 
End of Its 2010 Outburst}

\author{John A. Tomsick\altaffilmark{1},
Kazutaka Yamaoka\altaffilmark{2},
Stephane Corbel\altaffilmark{3},
Emrah Kalemci\altaffilmark{4},
Simone Migliari\altaffilmark{5}, and
Philip Kaaret\altaffilmark{6}}

\altaffiltext{1}{Space Sciences Laboratory, 7 Gauss Way, University of 
California, Berkeley, CA 94720-7450, USA (jtomsick@ssl.berkeley.edu)}

\altaffiltext{2}{Solar-Terrestrial Environment Laboratory, 
Department of Particles and Astronomy, Nagoya University,
Furocho, Chikusa-ku, Nagoya, Aichi 464-8601, Japan}

\altaffiltext{3}{AIM - Unit\'e Mixte de Recherche CEA - CNRS -
Universit\'e Paris VII - UMR 7158, CEA-Saclay, Service d'Astrophysique,
91191 Gif-sur-Yvette Cedex, France}

\altaffiltext{4}{Sabanci University, Orhanli - Tuzla, Istanbul, 
34956, Turkey}

\altaffiltext{5}{Department d'Astronomia i Meteorologia, Universitat
de Barcelona, Marti I Franques 1, 08028 Barcelona, Spain}

\altaffiltext{6}{Department of Physics and Astronomy,
University of Iowa, Van Allen Hall, Iowa City, IA 52242, USA}

\begin{abstract}

Here we report on {\em Swift} and {\em Suzaku} observations near the end of an 
outburst from the black hole transient 4U~1630--47 and {\em Chandra} observations 
when the source was in quiescence.  4U~1630--47 made a transition from a soft 
state to the hard state $\sim$50\,d after the main outburst ended.  During this 
unusual delay, the flux continued to drop, and one {\em Swift} measurement found 
the source with a soft spectrum at a 2--10\,keV luminosity of 
$L = 1.07\times 10^{35}$\,erg\,s$^{-1}$ for an estimated distance of 10\,kpc.  
While such transients usually make a transition to the hard state at 
$L/L_{\rm Edd} = 0.3$--3\%, where $L_{\rm Edd}$ is the Eddington luminosity, the 
4U~1630--47 spectrum remained soft at $L/L_{\rm Edd} = 0.008$\,$M_{10}^{-1}$\%
(as measured in the 2--10\,keV band), where $M_{10}$ is the mass of the black 
hole in units of 10\Msun.  An estimate of the luminosity in the broader
0.5--200\,keV bandpass gives $L/L_{\rm Edd} = 0.03$\,$M_{10}^{-1}$\%, which is
still an order of magnitude lower than typical.  We also measured an exponential 
decay of the X-ray flux in the hard state with an e-folding time of $3.39\pm 0.06$\,d, 
which is much less than previous measurements of 12--15\,d during decays by
4U~1630--47 in the soft state.  With the $\sim$100\,ks {\em Suzaku} observation, 
we do not see evidence for a reflection component, and the 90\% confidence limits 
on the equivalent width of a narrow iron K$\alpha$ emission line are $<$40\,eV
for a narrow line and $<$100\,eV for a line of any width, which is consistent
with a change of geometry (either a truncated accretion disk or a change in
the location of the hard X-ray source) in the hard state.  Finally, we report 
a 0.5--8\,keV luminosity upper limit of $<$$2\times 10^{32}$\,erg\,s$^{-1}$
in quiescence, which is the lowest value measured for 4U~1630--47 to date.

\end{abstract}


\section{Introduction}

Most of the binary black hole systems in our Galaxy are X-ray transients 
that undergo outbursts where the luminosity can change by factors of 
$10^{8}$ or $10^{9}$.  As the observational properties of these systems
change during outbursts that last for a few months to a few years, they
can undergo state transitions.  There are different schemes for classifying
states \citep[e.g.,][]{hb05}, but one useful classification depends largely 
on whether the thermal, soft power-law, or hard power-law component 
dominates the $\sim$1--20\,keV energy spectrum.  These are called the 
thermal-dominant, steep power-law, and hard states, respectively \citep{mr06}, 
and intermediate states also occur.  Physically, the thermal-dominant state 
occurs when the optically thick accretion disk extends near or all the way 
to the innermost stable circular orbit (ISCO), and this component is well-modeled 
with a multi-temperature disk component \citep{mcclintock06}.  While radio jets 
are quenched during the thermal-dominant state \citep{fender99,russell11}, the 
hard state has steady jets that emit in the radio, IR, and perhaps at higher
frequencies \citep{corbel00,fender01,cf02,russell13}.  

Most often, black hole transients have outbursts where the source rises in the
hard state, makes a transition to the thermal-dominant or steep power-law state
at a relatively high luminosity, and then transits back to the hard state at
lower luminosity \citep{fbg04,corbel04,belloni05,dunn10}.  This hysteresis effect 
and its cause have been a topic of extensive discussion, and many theories have 
been advanced to explain it.  One idea is that sources tend to stay in the hard 
state during the rise because the hard X-ray emission keeps the corona hot while 
sources tend to stay in the soft state (i.e., either the thermal-dominant 
or steep power-law state) during decay because the soft X-ray emission 
keeps the corona cool \citep{mlm05,liu05}.  Another possibility is the two-flow 
picture where there is a Keplerian flow that corresponds to the optically thick 
accretion disk and a sub-Keplerian flow with the energetic electrons that produce 
the hard X-rays \citep{ct95}.  In this model, the bright hard state can be
explained by a longer (viscous) time scale for changes in the Keplerian flow 
compared to the sub-Keplerian flow \citep{debnath13}.

In other potential explanations for the hysteresis, the large-scale magnetic field 
in the disk, which can depend on the accretion state or the type of accretion flow,
plays an important role \citep{petrucci08,ba14}.  The \cite{ba14} model explains the 
hysteresis by invoking the connection between the magnetic field and the viscosity.
With the magnetorotational instability being the source of viscosity \citep{bh91}, 
changes in magnetic field imply changes in disk viscosity, and hysteresis is produced 
because the transition luminosity is given by $L\sim \alpha^{2} L_{\rm Edd}$ \citep{emn97}, 
where $\alpha$ is the viscosity parameter and $L_{\rm Edd}$ is the Eddington luminosity.  
\cite{petrucci08} also emphasize the role of the magnetic field in producing the 
hysteresis, but this is primarily through the possible connection between the
magnetic field in the inner disk and the production of jets.  \cite{petrucci08}
and \cite{ba14} discuss how changes in the magnetic field may explain the the observed 
jet behaviors such as the steady jet in the hard state and the impulsive relativistic
jets that are typically produced when systems make a transition from a bright hard
state \citep{fbg04,corbel04,fhb09}.

In this context, 4U~1630--47 is an interesting source since the large number of
outbursts allows us to compare their properties.  At least one well-studied
outburst follows the typical hysteresis pattern.  In 1998, the source was
seen in the hard state during the rise \citep{dieters00}, radio jets were 
produced \citep{hjellming99}, and it made a transition back to the hard state
at a much lower luminosity than the transition during the rise \citep{tk00}.
However, other outbursts have shown very different behavior.  A detailed
comparison between the 1998 outburst and the 2002--2004 outburst showed that
the 2002--2004 outburst was much softer \citep{tomsick05c}, and it did not 
exhibit bright hard states \citep{tomsick05c,abe05} or radio emission 
\citep{hannikainen02}.  The 2010 outburst was very similar to 2002--2004 in 
terms of hardness and evolution (see Appendix); however, while there was no 
observational coverage of the end of the 2002--2004 outburst, we have obtained 
such coverage in 2010.

Given the fact that 4U~1630--47 is in a crowded region of the Galactic Plane, 
source confusion was a major issue for the {\em Rossi X-ray Timing Explorer}
\citep[{\em RXTE};][]{brs93} in following the source to low luminosities during 
previous outbursts \citep{tomsick05c}.  Thus, in this work, we have used X-ray 
imaging observations with {\em Swift}, {\em Suzaku}, and {\em Chandra} to study 
the end of the 2010 outburst and quiescence (although the {\em RXTE} measurements
are shown in the Appendix).  A main motivation for this study is to determine the 
time of the transition to the hard state and to obtain a long observation with 
{\em Suzaku} after the transition in order to measure the iron K$\alpha$ emission 
line that comes from the reflection component \citep[e.g.,][]{fabian89}.  This 
follows a similar campaign where we observed GX~339--4 and obtained iron line 
measurements that provided evidence for truncation of the accretion disk 
\citep{tomsick09c}.  In the following, we describe the observations and how the 
data were analyzed (Section 2), report the results of the analysis, focusing on 
the evolution of the energy spectrum (Section 3), and then provide a discussion 
of the evolution of the spectral states and the implications for the constraints 
on the iron line and the quiescent luminosity limit (Section 4).

\section{Observation and Data Reduction}

We report on {\em Swift} and {\em Suzaku} observations of 4U~1630--47 taken between 
2010 July 15 and 2010 Aug 17 near the end of an outburst and {\em Chandra} 
observations made in 2011 June when the source was in quiescence.  The Observation 
IDs, start and stop times, and total exposure times are listed in Table~\ref{tab:obs}, 
which lists four relatively short (690--4,617\,s) {\em Swift} observations, a long 
($\sim$100\,ks) {\em Suzaku} observation, and two 19--20\,ks {\em Chandra} 
observations.

We reduced the {\em Swift} and {\em Suzaku} data using High Energy
Astrophysics Software (HEASOFT) v6.15 along with the 2013 March 13
(for {\em Swift}) and 2013 September 16 (for {\em Suzaku}) releases 
of the Calibration Database (CALDB).  For {\em Swift}, we produced
photon event lists and exposure maps using {\ttfamily xrtpipeline}.
The X-ray Telescope \citep[XRT;][]{burrows05} was in photon counting 
mode, which provides 2-dimensional imaging information.  For the 
source spectrum, we extracted the photons from a circle with a 20-pixel 
($47^{\prime\prime}$) radius centered on 4U~1630--47.  The background 
spectrum came from a source-free region in another part of the field 
of view.  We used {\ttfamily xrtmkarf} to make an ancillary response 
matrix and included an exposure map correction.  For the response matrix, 
we used the file selected from the CALDB by the {\ttfamily xrtmkarf} 
program.

For {\em Suzaku}, we reprocessed the data from the X-ray Imaging
Spectrometers \citep[XISs;][]{koyama07} using {\ttfamily aepipeline}
in order to apply the most recent calibrations.  We made an event 
list for each of the three XIS units (XIS0, XIS1, and XIS3).  We
extracted light curves and spectra using a circular source region
with a radius of 100 pixels ($104^{\prime\prime}$).  For background
regions, we used a rectangular region close to the edge of the field 
of view with dimensions of $172^{\prime\prime}$ by $224^{\prime\prime}$.
We used {\ttfamily xisrmfgen} and {\ttfamily xissimarfgen} to make
the response files.  For the energy spectra, we combined XIS0 and
XIS3 as they have very similar response matrices.

We analyzed data from {\em Chandra} ObsIDs 12530 and 12533, which are
observations made as part of the Norma spiral arm survey (PI: Tomsick).  
The Advanced CCD Imaging Spectrometer \citep[ACIS;][]{garmire03} instrument
was used for both ObsIDs, and the aimpoint was on the ACIS-I detector.
4U~1630--47 was $3.\!^{\prime}4$ and $12.\!^{\prime}9$ from the aimpoint
for ObsIDs 12530 and 12533, respectively.  For data reduction, we used 
the {\em Chandra} Interactive Analysis of Observations (CIAO) software 
and made event lists using {\ttfamily chandra\_repro}.  An inspection of 
the 0.5--8\,keV images does not show any evidence for a source at the 
position of 4U~1630--47 for either observation. Given the large 
off-axis angle for ObsID 12533, the sensitivity is much better for
ObsID 12530, and we report on a detailed analysis of the data from
ObsID 12530 below.

\section{Results}

\begin{figure}
\includegraphics[scale=0.5]{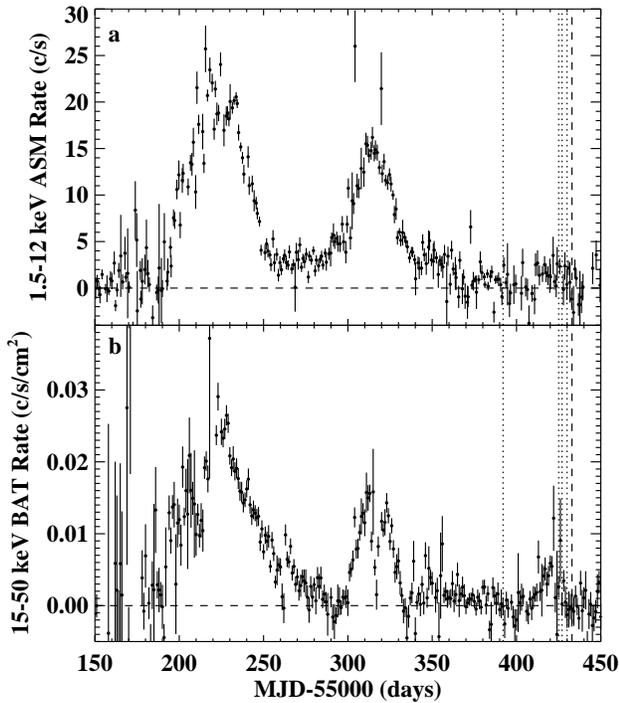}
\caption{\small X-ray light curves for 4U~1630--47 during the multi-peaked
2010 outburst, including {\em (a)} {\em RXTE}/ASM measurements in the 1.5--12\,keV 
band and {\em (b)} {\em Swift}/BAT measurements in the 15--50\,keV band.  The
times of the {\em Swift} (dotted lines) and {\em Suzaku} (vertical dashed line) 
observations are indicated.\label{fig:lc_with_bat}}
\end{figure}

\begin{figure}
\includegraphics[scale=0.5]{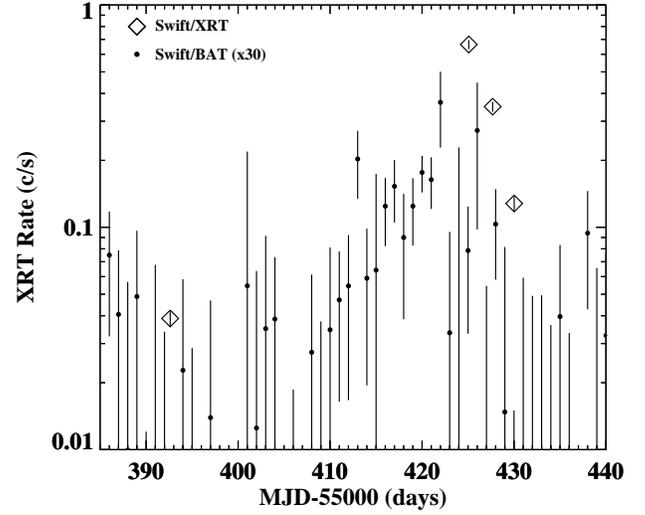}
\caption{\small {\em Swift}/XRT count rates for 4U~1630--47 in the 0.5--10\,keV band 
during the reflare near the end of the outburst.  The {\em Swift}/BAT 15--50\,keV
measurements are also shown.  The rates (in counts\,s$^{-1}$\,cm$^{-2}$) from 
Figure~\ref{fig:lc_with_bat}b have been multiplied by a factor of 30 for easier 
comparison to the XRT measurements.\label{fig:lc_xrt_bat}}
\end{figure}

Figure~\ref{fig:lc_with_bat}a shows the {\em RXTE} All-Sky Monitor (ASM) 
light curve for the 2010 outburst.  The {\em Swift} X-ray Telescope (XRT) 
and {\em Suzaku} observations occurred well after the bright part of the 
outburst, but the {\em Swift} Burst Alert Telescope (BAT) light curve 
(Figure~\ref{fig:lc_with_bat}b) shows that a reflare occurred during this 
time.  The reflare was also reported based on observations with the 
{\em RXTE} Proportional Counter Array \citep[PCA;][]{ty10}, and the increase 
seen by the PCA (see Appendix) led to the XRT observations.  The XRT 
observations confirm that the reflare is from 4U~1630--47 as reported in 
\cite{ty10} and as shown in Figure~\ref{fig:lc_xrt_bat}.

For spectral analysis, we used the XSPEC software package \citep{arnaud96}, 
fitting the four {\em Suzaku} spectrum and the {\em Swift} spectra in a 
uniform way.  Due to systematic uncertainties in the XIS response matrix 
associated with absorption edges in the instrument materials, we did not 
include the 1.65--1.9\,keV or 2.2--2.4\,keV energy bands, but we used the 
rest of the 0.4--9\,keV and 0.4--12\,keV bandpass for XIS1 and XIS0/3, 
respectively.  We rebinned the spectra to at least 100 counts per bin 
for XIS1 and to at least 200 counts per bin for XIS0/3.  For the 0.5--10\,keV 
XRT spectra, we rebinned so that each bin (except for the highest energy bin) 
has a detection at the 3-$\sigma$ level or higher.  When fitting, we minimized 
the W-statistic, which is a generalization of the Cash statistic \citep{cash79} 
for the case of non-zero background, and we used $\chi^{2}$ as the test statistic.

\begin{figure}
\includegraphics[scale=0.5]{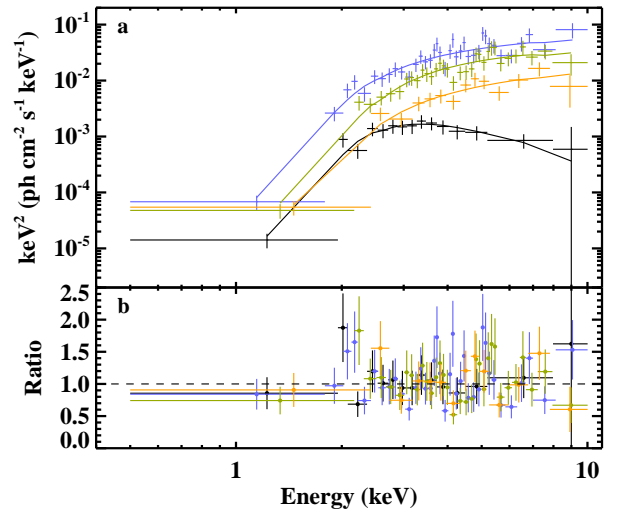}
\vspace{-1.2cm}
\caption{\small {\em (a)} The unfolded {\em Swift}/XRT spectra of 4U~1630--47
from observations \#1--4 (ObsIDs 00031224006-00031224009) fitted individually
with an absorbed power-law model (solid lines).  {\em (b)} The data-to-model
ratio residuals for power-law fits to the four spectra.  In both panels, 
observations \#1, \#2, \#3, and \#4 are plotted in black, yellow, green, 
and blue, respectively.\label{fig:swift_spectra}}
\end{figure}

We fitted each spectrum individually with an absorbed power-law model.
4U~1630--47 is known to have a high column density, and both interstellar
material as well as material local to the source may contribute.  As any 
local contribution could be variable, we left the column density as a free
parameter in our fits.  We used \cite{wam00} abundances and \cite{vern96} 
cross-sections for the absorption calculation.  The power-law model 
provides a good description of the {\em Swift} spectra as indicated by 
the reduced-$\chi^{2}$ ($\chi^{2}_{\nu}$) values reported in 
Table~\ref{tab:parameters}.  For observation \#1, the spectrum is very 
soft with a power-law photon index of $\Gamma = 4.3\pm 1.0$ (90\% confidence 
errors).  The spectrum hardened dramatically after observation \#1, and 
during observations \#2, \#3, and \#4, the range of best fit values is 
$\Gamma = 1.4$--1.7, and the 90\% confidence upper limits on $\Gamma$ are 
$<$2.0, $<$2.2, and $<$2.1, for the three observations, respectively.  The 
spectral transition is illustrated in Figure~\ref{fig:swift_spectra}.

The softness of the spectrum for observation \#1 suggests that it is
more likely dominated by thermal emission rather than a power-law.  We
fit the spectrum with an absorbed multi-temperature blackbody model 
\citep{mitsuda84}, commonly called the disk-blackbody model, and this also 
provides an acceptable fit with $\chi^{2}_{\nu} = 0.70$ for 12 degrees of freedom 
(dof).  The disk-blackbody fit returns a value of $kT_{\rm in} = 0.94^{+0.25}_{-0.17}$\,keV 
for the temperature at the inner edge of the accretion disk, which is typical 
for accreting black holes in the soft state.  The normalization we measure, 
$N_{\rm diskbb} = 0.7^{+1.9}_{-0.5}$, is related to the inner disk radius according 
to $N_{\rm diskbb} = (R_{\rm in,km}/d_{10})^{2}/\cos{i}$, where $R_{\rm in,km}$ 
is the inner radius in units of km, $d_{10}$ is the distance to the source
in units of 10\,kpc, and $i$ is the disk inclination.  Thus, $N_{\rm diskbb} < 2.6$ 
corresponds to $R_{\rm in} < 1.6$\,km ($d_{10}/\sqrt{\cos{i}}$).  Given that 
the distance to 4U~1630--47 is estimated to be near 10\,kpc and that the binary 
inclination is not extremely high (since the source is not eclipsing), such a
small inner radius would only be consistent with a very low mass black hole:
$\sim$1--2\,\Msun~for the maximally rotating case and an even smaller mass
for slower rotation rates.

A much more likely scenario is that the observation \#1 spectrum actually has
two components: the disk-blackbody and a power-law.  We re-fitted spectrum \#1 
with such a two component model, and this allows for a significantly larger 
normalization $N_{\rm diskbb} < 197$ for a disk-blackbody component with 
$kT_{\rm in} < 1.1$\,keV (both are 90\% confidence upper limits).  With the
possibility of the higher $N_{\rm diskbb}$, the implied inner radius is 
$R_{\rm in} < 14$\,km ($d_{10}/\sqrt{\cos{i}}$), which is physically
reasonable.  While the $R_{\rm in}$ estimates or limits are useful for determining 
which spectral model is the most likely, it should be noted that, in addition to 
the distance and inclination uncertainties, there are other corrections that would 
be necessary \citep{mcclintock06} if our intent was to report a measurement of 
$R_{\rm in}$.  

\begin{figure}
\includegraphics[scale=0.5]{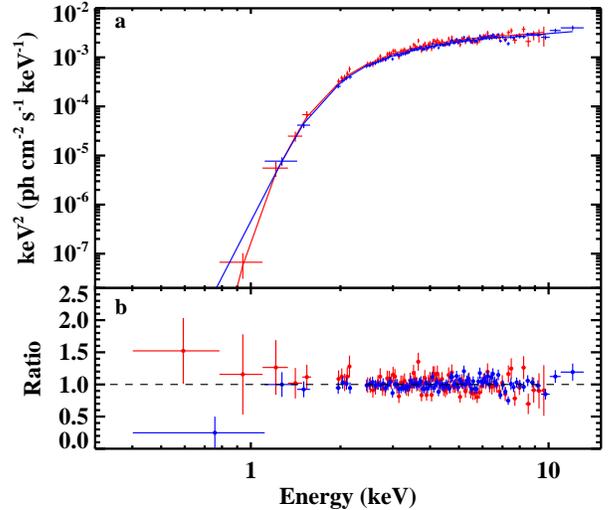}
\vspace{-1.2cm}
\caption{\small {\em (a)} The unfolded {\em Suzaku} spectrum of 4U~1630--47 
from observation \#5.  The blue points show the spectrum measured with the 
XIS0 and XIS3 units, and the red points show XIS1 (and these colors are used 
for both panels). {\em (b)} The data-to-model ratio residuals for an absorbed 
power-law model.\label{fig:spectrum_suzaku}}
\end{figure}

The {\em Suzaku}/XIS spectrum (observation \#5) is well-described by 
an absorbed power-law model (see Figure~\ref{fig:spectrum_suzaku} for 
the spectrum and residuals and Table~\ref{tab:parameters} for the 
parameter values) with $N_{\rm H} = (5.8\pm 0.3)\times 10^{22}$\,cm$^{-2}$ 
and $\Gamma = 1.58\pm 0.05$.  Such a value for the power-law photon 
index is expected for a black hole in the hard state.  We might also 
expect to see evidence for a reflection component in the spectrum.  
While the limited bandpass of XIS would not allow us to detect the 
Compton bump, which peaks near 20--40\,keV \citep{lw88}, the characteristic 
iron emission line and absorption edge is covered by XIS.  Although the 
residuals (see Figure~\ref{fig:spectrum_suzaku}b) do not show strong 
evidence for iron features, adding a Gaussian with a centroid of 
$E_{\rm line} = 6.3\pm 0.2$\,keV and a width of 
$\sigma_{\rm line} = 0.23^{+0.15}_{-0.12}$\,keV improves the fit 
from $\chi^{2}/\nu = 183.5/185$ to $\chi^{2}/\nu = 173.9/182$.  We used 
the XSPEC script {\ttfamily simftest} to produce 1000 simulated XIS 
spectra with an absorbed power-law as the input spectrum and to fit 
them with and without a Gaussian.  While fitting the simulated
spectra, $E_{\rm line}$ was restricted to 6--7\,keV and $\sigma_{\rm line}$
was kept within the 90\% confidence error range found when fitting
the actual data.  We found improvements in the fit as large as
the observed improvement for 15 of the simulated spectra, indicating 
that the significance of the line is 2.2-$\sigma$, which we do not
consider to be a significant detection.  With the Gaussian parameters
free, the 90\% confidence upper limit on the equivalent width ($EW$) 
of the line is $<$100\,eV.  For a narrow line at 6.4\,keV, the value
is $EW < 41$\,eV.

To understand the implications of these upper limits for reflection
models, we fit the XIS spectra with a model consisting of an 
absorbed power-law model and the {\ttfamily reflionx} reflection
model \citep{rf05}.  When leaving the parameters for this model
free, the best fit value for the ionization parameter is its 
minimum value ($\xi = 10$\,erg\,cm\,s$^{-1}$) in order to fit
the low-level emission feature near 6.4\,keV.  The iron abundance
is not constrained, and we fix it to the solar value 
(1.0 in the {\ttfamily reflionx} model).  Fitting the spectrum
with this model gives a 90\% confidence upper limit on the reflection 
covering fraction of $\Omega/2\pi < 0.11$.  While this upper limit
is valid for the case where there is no relativistic smearing of
the reflection component (e.g., if the disk is truncated), if we
allow the reflection component to be smeared using, e.g., 
{\ttfamily kdblur} \citep{laor91}, a significantly higher 
covering fraction is possible.

\begin{figure}
\includegraphics[scale=0.5]{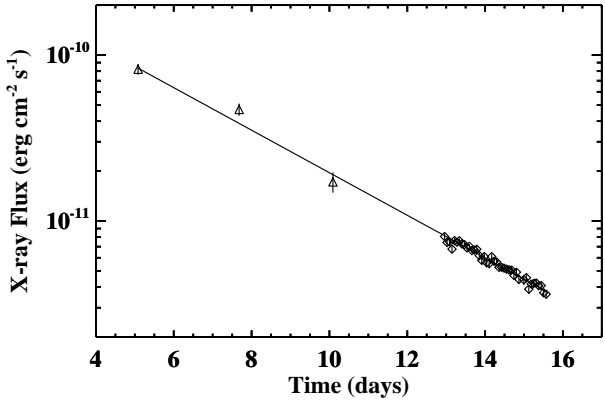}
\caption{\small The 1--10\,keV unabsorbed fluxes for 4U~1630--47
from {\em Swift} observations \#2--4 (triangles) and the {\em Suzaku} 
observation (diamonds).  For the time axis, zero corresponds to MJD~55,420.
The line is a fit with an exponential with an e-folding decay time of
$3.39\pm 0.06$\,d.\label{fig:lc_decay}}
\end{figure}

The evolution of the source during observations \#1--5 is consistent 
with a transition from a low-luminosity soft or intermediate state 
during observation \#1 to a hard state in observation \#2.  We 
investigated the source evolution during the decay by making a light 
curve of the absorbed 1--10\,keV fluxes during observations \#2--5.  
We calculated one flux point for each of the {\em Swift} observations 
and made a light curve with 5500\,s (approximately the satellite 
orbital period) time resolution for {\em Suzaku}.  We converted the 
XIS count rates to fluxes using the power-law fit to the energy spectrum.  
The model shown in Figure~\ref{fig:lc_decay} shows that an exponential 
with an e-folding decay time of $3.39\pm 0.06$\,d provides a good 
description of the {\em Suzaku} and {\em Swift} points 
($\chi^{2}_{\nu} = 1.02$ for 42 dof).

Although there is a long gap of nearly 10 months between the 
{\em Suzaku} and {\em Chandra} observations, the {\em Chandra} observation
provides an opportunity to determine if the source continued its decay
into quiescence.  For the observation on 2011 June 16 (ObsID 12530), 
4U~1630--47 was $3.\!^{\prime}4$ from the {\em Chandra} aimpoint, where 
the 90\% encircled energy fraction (EEF) radius (for 4.5\,keV photons) is
$2.\!^{\prime\prime}9$ as determined from the {\em Chandra} PSF 
Viewer\footnote{http://cxc.harvard.edu/cgi-bin/prop\_viewer/build\_viewer.cgi?psf}.
From previous radio detections, the 4U~1630--47 position is known to 
subarcsecond accuracy \citep{hjellming99}, and we find only one photon
(with an energy of 1.48\,keV) within $2.\!^{\prime\prime}9$ of that position 
during the 19,260\,s {\em Chandra} observation.  The prediction for the 
number of 0.5--8\,keV background counts in the same region is 0.57 counts.
Thus, using Poisson statistics, the upper limit on the count rate is 
$<$$1.7\times 10^{-4}$\,s$^{-1}$.  Assuming that the spectrum has a power-law
shape with $N_{\rm H} = 6\times 10^{22}$\,cm$^{-2}$ and $\Gamma = 2$
\citep{plotkin13,reynolds14}, the upper limits on the absorbed and unabsorbed 
0.5--8\,keV fluxes are $<$$4.7\times 10^{-15}$\,erg\,cm$^{-2}$\,s$^{-1}$ and 
$<$$1.6\times 10^{-14}$\,erg\,cm$^{-2}$\,s$^{-1}$.  

\section{Discussion}

Due to its location in a crowded region of the Galactic plane, imaging X-ray 
observations are required to reliably follow the evolution and determine 
the properties of 4U~1630--47 at low luminosities.  We have used {\em Swift}
and {\em Suzaku} observations during the decay phase of the 2010 outburst and
a {\em Chandra} observation in quiescence to provide new information about 
4U~1630--47 at low luminosities.  In the following, we discuss these findings
in the order they were observed: the soft state, the transition to the hard
state and subsequent evolution, the limits on a reflection component, and the
limits on the quiescent luminosity.

\subsection{A Soft State at Low Luminosities}

During {\em Swift} observation \#1, the spectrum of 4U~1630--47 was very soft, 
and it was likely dominated by thermal emission.  Despite the low flux, the 
source had clearly not made a transition to the hard state, which is notable
because black hole transients typically make a transition to the hard state 
during outburst decay at luminosities between 0.3\% and 3\% of the Eddington
limit \citep{maccarone03,kalemci13}.  For 4U~1630--47, the unabsorbed 2--10\,keV 
flux was $9\times 10^{-12}$\,erg\,cm$^{-2}$\,s$^{-1}$ during this observation.
The conversion to an Eddington-scaled luminosity depends on the distance to
the system, the mass of the black hole, and a bolometric correction.  We follow 
previous work \citep[e.g.,][]{psw86} by assuming a distance of 10\,kpc, giving 
a 2--10\,keV luminosity of $1.07\times 10^{35}$\,erg\,s$^{-1}$.  Although the 
distance is somewhat uncertain, \cite{akv01} argue that it is less than 11\,kpc
since there is a Giant Molecular Cloud (GMC) at this distance \citep{corbel99a} 
in the direction of 4U~1630--47.  The measured column density for 4U~1630--47
can be as low as $5\times 10^{22}$\,cm$^{-2}$, but $N_{\rm H}$ would always be 
significantly larger if the source was behind the GMC
\citep[see][and references therein]{akv01}.  The black hole mass has not been 
measured, and we use a fiducial value of 10\Msun~for this calculation, giving 
an Eddington luminosity of $L_{\rm Edd} = 1.3\times 10^{39}$$M_{10}$\,erg\,s$^{-1}$, 
where $M_{10}$ is the mass of the black hole in units of 10\Msun.  Thus, in the 
2--10\,keV band, the Eddington-scaled luminosity is 0.008\,$M_{10}^{-1}$\%.  

With the limited bandpass of XRT, making the bolometric correction is not 
straightforward.  While it might be reasonable to extrapolate the steep 
($\Gamma = 4.3$) power-law to $>$10\,keV, extrapolating to lower energies
is not consistent with the fact that the origin of the soft X-ray emission
is likely thermal.  In Section 3, we fit the observation \#1 
spectrum with a disk-blackbody plus power-law model, and here, we use those 
best fit parameters for the extrapolation.  With that model, we found 
$N_{\rm H} = 10^{23}$\,cm$^{-2}$, $kT_{\rm in} = 0.55$\,keV, $N_{\rm diskbb} = 12$, 
$\Gamma = 2.6$, and an unabsorbed 2--10\,keV flux for the power-law
component of $3.0\times 10^{-12}$\,erg\,cm$^{-2}$\,s$^{-1}$.  In the 
0.5--200\,keV band, the unabsorbed flux is $3.0\times 10^{-11}$\,erg\,cm$^{-2}$\,s$^{-1}$, 
corresponding to $L/L_{\rm Edd} = 0.03$\,$M_{10}^{-1}$\%.  Thus, this detection 
of a soft state occurred at a luminosity that is at least an order of magnitude 
lower than the level where a transition to the hard state would be expected.

As a consistency check on the 4U~1630--47 distance and black hole mass, we
consider the highest fluxes seen from this source, which were 0.84 Crab
(1.5--12\,keV) during the 2002--2004 outburst \citep{tomsick05c} and 1.4 Crab 
(3--6\,keV) in 1977 \citep{csl97}.  From the 2002--2004 outburst, the highest
broadband (3--200\,keV) flux quoted was $3.9\times 10^{-8}$\,erg\,cm$^{-2}$\,s$^{-1}$, 
and considering the bolometric correction and the brighter 1977 outburst, the
maximum flux from 4U~1630--47 is at least $6\times 10^{-8}$\,erg\,cm$^{-2}$\,s$^{-1}$.
For a distance of 10\,kpc, this flux corresponds to $L/L_{\rm Edd} = 55$\,$M_{10}^{-1}$\%, 
and $L = L_{\rm Edd}$ would occur at a distance of 14\,kpc.  Although this limit
is less constraining than the $<$11\,kpc discussed above, this provides a second
line of evidence that the distance could not be very much more than the 10\,kpc
that we have assumed (unless the 4U~1630--47 black hole is more massive than 
other systems).  While we note that the distance could be less than 10\,kpc if
a significant fraction of the column density is local to the source, a lower
distance would make the luminosity when we observe the soft state even lower
and more unusual.

There are other cases of soft states at low luminosities, but the best known
examples are not at a level as low as we are seeing for 4U~1630--47.  In the
detailed study of several black hole systems by \cite{kalemci13}, the lowest
luminosity soft state was seen at 0.3\% $L_{\rm Edd}$ for XTE~J1720--330.
1E~1740.7--2942 and GRS~1758--258 both showed unusual behavior by making a 
transition to a true thermal-dominant state (i.e., the spectra required a thermal 
disk component) after a drop in luminosity.  However, in both of these cases, the 
thermal-dominant state was seen at 1--2\% $L_{\rm Edd}$ \citep{smith01,delsanto05}.  
Even so, this might provide an interesting comparison to 4U~1630--47 since this 
low-soft state has been called a ``dynamical'' soft state \citep{smith07} based 
on the two-flow picture and the idea that there is some level of independence 
between the optically thick flow and the sub-Keplerian flow \citep{ct95}.

If the luminosity threshold for the transition to the hard state is set by
the magnetic field and viscosity in the disk \citep{petrucci08,ba14}, then
the interpretation for a low-soft state would be a weaker than usual 
large-scale magnetic field in the disk.  In the \cite{petrucci08} picture, 
the hard state jet-emitting disk is established when the disk magnetization, 
$\mu$, which is related to the disk magnetic field as well as the total
pressure (gas plus radiation) in the disk, reaches a specific threshold level 
\citep[called $\mu_{\rm max}$ in][]{petrucci08}.  Thus, a delayed transition to 
the hard state would be predicted if $\mu$ starts at a lower value or if it 
rises slowly.  As the rise of $\mu$ is primarily due to a drop in mass 
accretion rate, which we assume occurs during the decay of all outbursts, it 
is more likely that $\mu$ starts at a lower value.

As discussed in \cite{petrucci08} and shown in \cite{ba14}, a low magnetic 
field leads to a lower viscosity parameter, and a lower transition luminosity 
would be predicted, which is consistent with what we see in 4U~1630--47.  
If the magnetic field was lower than typical during the entire outburst, 
then this might also explain why we never observed a bright hard state 
(see Appendix) and also why radio emission was not reported during the 2010 
outburst or during the other outburst (2002--2004) which had very similar 
evolution in the hardness intensity diagram.

\subsection{Evolution in the Hard State}

\begin{figure}
\includegraphics[scale=0.5]{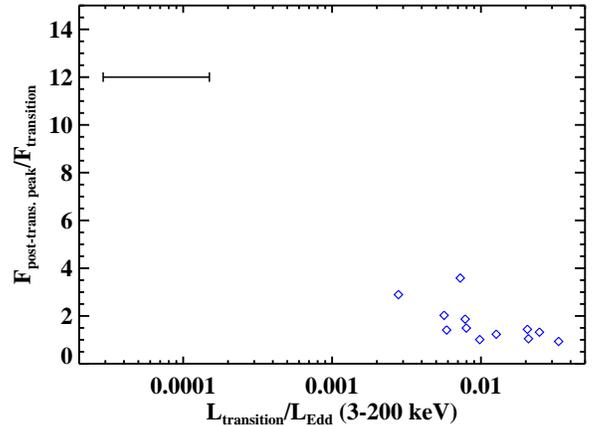}
\caption{\small Ratio of the peak X-ray flux after the transition to the hard
state ($F_{\rm post-trans.~peak}$) to the minimum flux prior to the transition
($F_{\rm transition}$) as a function of the Eddington luminosity fraction 
prior to the transition.  The blue diamonds are for the seven sources
and 12 outbursts reported in \cite{kalemci13}, where a 3--200\,keV band
was used.  The black error bar is for 4U~1630--47 during the 2010 transition
using a bandpass of 3--200\,keV for a distance of 10\,kpc.  The error region
corresponds to a black hole mass in a 3--15\Msun~range.\label{fig:increase}}
\end{figure}

When the transition to the hard state did finally occur, there was a very dramatic 
increase in the 2--10\,keV flux with an increase by at least a factor of 12 (see 
Figure~\ref{fig:lc_xrt_bat} and Table~\ref{tab:parameters}).  While increases in 
flux have been seen at the soft-to-hard state transition, they are more modest.
For XTE~J1650--500, the flux increased by a factor of 3--4 in the 3--20\,keV band 
\citep{kalemci03a}, and XTE~J1752--223 showed an increase by a factor of 3.5 in the 
3--25\,keV band \citep{chun13}.  Figure~\ref{fig:increase} shows the increase as a 
ratio of the peak X-ray flux after the transition to the flux prior to the 
transition ($F_{\rm post-trans.~peak}/F_{\rm transition}$) as a function of the 
Eddington fraction prior to the transition ($L_{\rm transition}/L_{\rm Edd}$)
for 4U~1630--47 compared to the other black hole transient outbursts studied by 
\cite{kalemci13}.  The values shown in Figure~\ref{fig:increase} are for the
3--200\,keV band, which is the bandpass used in the \cite{kalemci13} analysis.  

We suspect that there is a connection between the low luminosity soft state and 
this unusually large increase in the hard state.  One piece of support for this is 
that, in the hard state, the flux reached a high enough level to be at a typical 
level for a soft-to-hard state transition.  Based on the observation \#2 spectrum, 
and extrapolating the power-law to higher energies (i.e., making the bolometric
correction), we find that the 0.5--200\,keV unabsorbed flux is 
$7.3\times 10^{-10}$\,erg\,cm$^{-2}$\,s$^{-1}$, which corresponds to 
0.7\,$M_{10}^{-1}$\%.  In other words, if this hard state flare had happened 
$\sim$50\,d earlier, we would not have seen the low luminosity soft state,
nor the large increase in flux at the state transition, and the evolution would
have appeared to be typical.  Although this effect has not been seen to such
a large level in other sources, Figure~\ref{fig:increase} suggests that the
outbursts with the lowest luminosity transitions may tend to have larger flux
increases.

We also obtained a very good measurement
of the decay in the hard state (see Figure~\ref{fig:lc_decay}), and it is simply
an exponential with an e-folding time of $3.39\pm 0.06$\,d.  This number is much
shorter than previous e-folding decay times measured for 4U~1630--47.  During the
1996, 1998, and 1999 outbursts, the decay time scales for the 1.5--12\,keV light
curve were between 12 and 15\,d \citep{kuulkers97,tk00,abraham01}.  However, an 
important difference is that the measurements made in the 1990s were for the part 
of the outburst when the source was still in the soft state.  Within the two-flow 
picture, this suggests the possibility that the different e-folding times are 
related to different viscous time scales for the two flows.  The short time scale
could also be related to the fact that the transition occurred at a very low
mass accretion rate so that the amount of material in the hard X-ray emitting 
region was more quickly depleted.

\subsection{Limits on a Reflection Component}

A main goal of the {\em Suzaku} observation was to place constraints on the
reflection component, which depends on the geometry of the accretion disk
and the hard X-ray source.  With XIS, the main constraint comes from a
measurement of the strength of the iron line.  During the brighter phases 
of outbursts, the Fe K$\alpha$ region of the 4U~1630--47 spectrum is complex, 
showing many different types of absorption and emission lines.  The absorption 
lines that have been detected are narrow \citep{kubota07} and are caused by 
material in a disk wind \citep{king14} or the accretion disk atmosphere 
\citep{rozanska14} or both.  Emission lines have also been detected with 
relatively narrow features possibly originating in the jets 
(Cui, Chen \& Zhang 2000; D{\'{\i}}az Trigo et al. 2013; but also
see Neilsen et al. 2014)\nocite{cui00a,diaztrigo13,neilsen14} and a broad 
line that is related to the reflection component \citep{king14}.  While it
is clear that lines are detected in the bright parts of the outburst, the 
question of whether there are iron features in the hard state is still open.  
Although a line was very clearly detected in the hard state at the end of the 
1998 outburst by {\em RXTE} \citep{tk00}, {\em INTEGRAL} imaging of the region 
found nearby sources \citep{tomsick05c} that could have at least contributed 
to this line.

An emission line from a reflection component was seen in GX~339--4 in the
hard state at a low luminosity of 0.14\% $L_{\rm Edd}$, and the narrow width 
and decreasing strength of the line \citep{tomsick09c} clearly shows a change 
in the system geometry compared to the broad line seen when the source was 
bright \citep{miller08}.  \cite{tomsick09c} interpreted this as truncation 
of the inner accretion disk, but \cite{dauser13} point out that this could 
also be due to a change in the geometry of the hard X-ray source with one 
possible scenario being that this emission comes from the jet.  For 4U~1630--47, 
our main result is an upper limit on the equivalent width of a narrow line 
at 6.4\,keV of $EW < 41$\,eV.  This limit is lower than the measured $EW$
of the line in GX~339--4, which was $77^{+12}_{-10}$\,eV.  The low $EW$ is 
consistent with 4U~1630--47 showing a continuation of the evolution seen for 
GX~339--4 with the disk becoming even more truncated or with the hard X-ray 
source moving even farther away from the disk (perhaps along the jet) as
the luminosity drops.  The 1-100\,keV Eddington-scaled luminosity of 
4U~1630--47 during the {\em Suzaku} observation was 0.03\,$M_{10}^{-1}$\%, 
which is about five times lower than for GX~339--4.  While we cannot rule 
out the possibility of a broad line, even that limit ($EW < 100$\,eV) is 
lower than the limit of 180\,eV suggested by \cite{miller07} as being a 
``significant non-detection'' in the sense that a line from a disk that extends 
to the innermost stable circular orbit is predicted to have a line with an
$EW$ near 180\,eV.

\subsection{New Limit on Quiescent Luminosity}

There has been significant interest in the quiescent luminosities of stellar
mass black holes and the comparison to neutron stars \citep{menou99,garcia01}.  
The typically lower black hole luminosities have been taken as evidence for 
the existence of a black hole event horizons as accretion energy can be 
advected across an event horizon but not through a neutron star surface
\citep[e.g.,][]{ngm97}; although, there may be other possibilities
for where the accretion energy goes, such as into the jet \citep{fgj03}.
Thus, it is somewhat surprising that a source as well-studied as 4U~1630--47
does not have a sensitive measurement of the quiescent flux.  Assuming a
distance of 10\,kpc, the lowest previously reported luminosity came from
a {\em ROSAT} measurement in the 0.2--2.4\,keV band that was made in 1992 
\citep{paw95}, and the value was $<$$7\times 10^{32}$\,erg\,s$^{-1}$
\citep{parmar97}.  Given the high column density of 4U~1630--47 and the
very soft {\em ROSAT} bandpass, this limit is highly dependent on the 
spectral shape assumed.  Our 0.5--8\,keV unabsorbed flux limit from 
{\em Chandra} corresponds to a luminosity of $<$$2\times 10^{32}$\,erg\,s$^{-1}$.
With the lower value and the wider bandpass, this limit is significantly
more constraining than the previous limit.

A value of $2\times 10^{32}$\,erg\,s$^{-1}$ would be the third brightest 
quiescent black hole out of 15 black hole systems with reported measurements
\citep{nm08,rm11}.  The luminosity level that divides neutron stars and black 
holes depends on the orbital period of the binary since the quiescent mass 
accretion rate is expected to be correlated with orbital period.  The limit 
we measure for 4U~1630--47 would fall in the black hole region for orbital periods 
higher than $\sim$20\,hr.  Although the orbital period is not known for 
4U~1630--47, a period near 20\,hr would not be surprising.  In the future, if the 
orbital period is found to be lower, this would motivate deeper quiescent
X-ray observations to try to obtain a measurement of the quiescent luminosity.

\section{Conclusions}

X-ray observations of 4U~1630--47 at the end of its 2010 outburst show that
the source decayed to a much lower luminosity than is typical before making 
a transition to the hard state.  About two weeks before the transition to the
hard state, the source had a soft and likely thermal X-ray spectrum and the 
0.5--200\,keV luminosity was $\sim$$3.5\times 10^{35}$\,erg\,s$^{-1}$ (assuming
a distance of 10\,kpc).  This corresponds to $L/L_{\rm Edd} = 0.03$\,$M_{10}^{-1}$\%, 
which is at least an order of magnitude lower than typical transition luminosities.
We discuss this evolution in terms of theoretical models for hysteresis of black
hole state transitions, and based on work by \cite{petrucci08} and \cite{ba14},
we suggest that the behavior could be explained by a lower than typical large-scale
magnetic field in the accretion disk.  We also consider the two-flow model of
\cite{ct95} as the evolution may suggest decoupling between the optically thick 
Keplerian disk and the sub-Keplerian flow.

Any conclusion about hysteresis also needs to consider the geometry of the system, 
including the location of the inner radius of the optically thick disk and whether 
there is significant X-ray emission from the jet in the hard state.  With {\em Suzaku}, 
we have placed tight limits on the presence of a reflection component from 4U~1630--47 
in the hard state at $L/L_{\rm Edd} = 0.03$\,$M_{10}^{-1}$\%, and the lack of a reflection 
component is consistent with a large inner radius.  More detailed theoretical work is 
required to determine whether a non-detection of reflection is consistent with a small 
inner radius and an increased height for the hard X-ray source above the disk.  Finally, 
we report a significantly lower upper limit on the quiescent X-ray luminosity, which is 
interesting to compare to the other measurements of quiescent luminosities of neutron 
star and black hole transients.

\acknowledgments

Partial support for this work was provided by NASA through {\em Suzaku} Guest
Observer grant NNX11AC89G, {\em Swift} Guest Observer grants NNX10AK36G and 
NNX13AJ81G, and {\em Chandra} Guest Observer grant GO1-12068A.  SC acknowledges
funding support from the French Research National Agency: CHAOS project
ANR-12-BS05-0009 (http://www.chaos-project.fr).  EK thanks TUBITAK for support
under grant 111T222.  JAT acknowledges useful discussions with D.~Smith.  
EK acknowledges useful discussions with M.~Begelman and P.-O.~Petrucci.

\appendix
\begin{center}
{\bf\normalsize Appendix}
\end{center}

\begin{figure}
\includegraphics[scale=1.0]{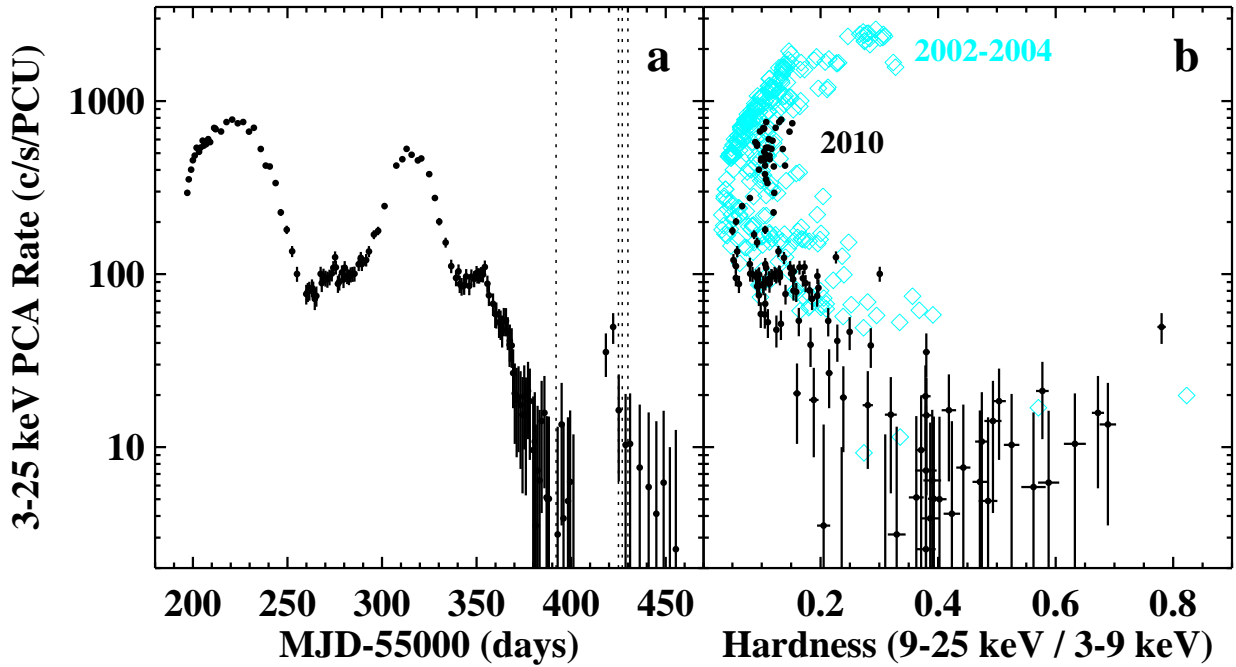}
\caption{\small {\em (a)} {\em RXTE}/PCA count rates for 4U~1630--47 
during the 2010 outburst (using PCU2 only) in the 3--25\,keV band. 
Estimates for instrumental background and the Galactic Ridge emission
(7.1 c/s) are subtracted off.  In addition, a systematic uncertainty
of $\pm$10 c/s is included.  The vertical dotted lines mark the times 
of the {\em Swift} observations.  {\em (b)} {\em RXTE}/PCA hardness-intensity 
diagram.  We do not include any systematic errors on the hardness measurements, 
and we estimate that the systematic uncertainty becomes significant below
$\sim$100 c/s.  For comparison, the values for the 2002--2004 outburst 
\citep[from][]{tomsick05c} are shown with cyan diamonds.
\label{fig:lc_hi_pca}}
\end{figure}

During the 2010 outburst, $\sim$150 pointed observations of 4U~1630--47 were made 
with the {\em Rossi X-ray Timing Explorer}.  These observations occurred between 
January 1 and September 16 and have ObsIDs starting with 95360-09 and 95702-03.  
The exposure times range from 500\,s to 4100\,s per observation with the average
being close to 2000\,s.  We used HEASOFT to extract energy spectra from the
Proportional Counter Array (PCA) for each of the observations and also to determine 
the PCA count rate in the 3--9\,keV (Channels 3--17) and 9--25\,keV (Channels
18--52) energy bands.  Figure~\ref{fig:lc_hi_pca} shows the light curve and
the hardness-intensity diagram for all of the observations.  The hardness-intensity
diagram for the 2010 outburst is similar to the one seen during the 2002--2004
outburst \citep{tomsick05c}, and Figure~\ref{fig:lc_hi_pca}b shows a direct
comparison.  The 2002--2004 and 2010 outbursts both have very different evolution 
in the hardness-intensity diagram compared to the typical ``q-shaped'' evolution 
for black hole transients \citep{dunn10}, which 4U~1630--47 showed during its 1998 
outburst (Tomsick et al. 2005 shows a comparison between the 1998 and the 2002--2004 
outbursts)\nocite{tomsick05c}.

At the lower count rates, the PCA measurements become suspect due to the fact that
the PCA has a collimated field of view (FOV) with a radius of $1^{\circ}$ (full-width 
zero intensity).  It is known that IGR~J16320--4751 is a persistent and highly variable
X-ray source that is in the FOV \citep{tomsick05c}, and there are also contributions from
other sources (i.e., Galactic Ridge emission).  Here, we use contemporaneous {\em RXTE} 
and {\em Swift} observations to estimate the possible contribution from other sources.  
For {\em Swift} observations \#1--4, there are {\em RXTE} observations within 7.2, 1.2, 
25.8, and 26.8\,h, respectively (see Table~\ref{tab:confusion}).  From observation \#1, 
it is clear that the level of contamination is severe.  The best fit absorbed power-law
model measured during {\em Swift} observation \#1 predicts a 3--25\,keV PCA count rate
of 0.23 c/s, but the measured rate is 45 times higher.  After accounting for the 
exponential decay from 4U~1630--47, the actual rates for observations \#2--4 are, 
respectively, 2.1, 3.5, and 8.8 times higher than the predicted rates.    In terms of 
count rates, the actual rates are 10.1, 12.1, 12.5, and 15.6 c/s higher than the predicted 
rates for observations \#1--4.  Although the division between contamination by variable 
and steady sources is not certain, the typical total contamination is $\sim$10--16 c/s, 
and the variable contribution (perhaps dominated by IGR~J16320--4751) is $>$5 c/s.

The final PCA rates shown in Figure~\ref{fig:lc_hi_pca} are corrected for Galactic
Ridge emission by subtracting off the rate measured during the observation with
the lowest count rate, which is 7.2 c/s.  Furthermore, we estimate that the 
systematic contribution to the count rate uncertainty due to variable sources (such 
as IGR~J16320--4751) is, conservatively, $\pm$10 c/s, and we include this uncertainty
in the error bars shown in Figure~\ref{fig:lc_hi_pca}.  However, it should be noted
that no systematic uncertainty is included in the hardness (Figure~\ref{fig:lc_hi_pca}b).
Thus, while the values plotted are good representations of the rates from 4U~1630--47, 
the hardnesses for the points with count rates below $\sim$100 c/s are subject to
considerable uncertainty.

\clearpage


\begin{thebibliography}{}

\bibitem[\protect\astroncite{{Abe} et~al.}{2005}]{abe05}
{Abe}, Y., {Fukazawa}, Y., {Kubota}, A., {Kasama}, D., \& {Makishima}, K.,
  2005, PASJ, 57, 629

\bibitem[\protect\astroncite{{Abraham} et~al.}{2001}]{abraham01}
{Abraham}, L., {Agrawal}, V., {Sreekumar}, P., \& {Seetha}, S.,  2001, Bulletin
  of the Astronomical Society of India, 29, 365

\bibitem[\protect\astroncite{{Arnaud}}{1996}]{arnaud96}
{Arnaud}, K.~A.,  1996,
\newblock in Astronomical Data Analysis Software and Systems V, ed. G.~H.
  {Jacoby}, J. {Barnes}, Vol. 101, ~17

\bibitem[\protect\astroncite{{Augusteijn}, {Kuulkers} \& {van
  Kerkwijk}}{2001}]{akv01}
{Augusteijn}, T., {Kuulkers}, E., \& {van Kerkwijk}, M.~H.,  2001, A\&A, 375,
  447

\bibitem[\protect\astroncite{{Balbus} \& {Hawley}}{1991}]{bh91}
{Balbus}, S.~A., \& {Hawley}, J.~F.,  1991, ApJ, 376, 214

\bibitem[\protect\astroncite{{Begelman} \& {Armitage}}{2014}]{ba14}
{Begelman}, M.~C., \& {Armitage}, P.~J.,  2014, ApJ, 782, L18

\bibitem[\protect\astroncite{{Belloni} et~al.}{2005}]{belloni05}
{Belloni}, T., {Homan}, J., {Casella}, P., {van der Klis}, M., {Nespoli}, E.,
  {Lewin}, W.~H.~G., {Miller}, J.~M., \& {M{\'e}ndez}, M.,  2005, A\&A, 440,
  207

\bibitem[\protect\astroncite{{Bradt}, {Rothschild} \& {Swank}}{1993}]{brs93}
{Bradt}, H.~V., {Rothschild}, R.~E., \& {Swank}, J.~H.,  1993, A\&AS, 97, 355

\bibitem[\protect\astroncite{{Burrows} et~al.}{2005}]{burrows05}
{Burrows}, D.~N., et~al., 2005, Space Science Reviews, 120, 165

\bibitem[\protect\astroncite{{Cash}}{1979}]{cash79}
{Cash}, W.,  1979, ApJ, 228, 939

\bibitem[\protect\astroncite{{Chakrabarti} \& {Titarchuk}}{1995}]{ct95}
{Chakrabarti}, S., \& {Titarchuk}, L.~G.,  1995, ApJ, 455, 623

\bibitem[\protect\astroncite{{Chen}, {Shrader} \& {Livio}}{1997}]{csl97}
{Chen}, W., {Shrader}, C.~R., \& {Livio}, M.,  1997, ApJ, 491, 312

\bibitem[\protect\astroncite{{Chun} et~al.}{2013}]{chun13}
{Chun}, Y.~Y., et~al., 2013, ApJ, 770, 10

\bibitem[\protect\astroncite{{Corbel} et~al.}{1999}]{corbel99a}
{Corbel}, S., {Chapuis}, C., {Dame}, T.~M., \& {Durouchoux}, P.,  1999, ApJ,
  526, L29

\bibitem[\protect\astroncite{{Corbel} \& {Fender}}{2002}]{cf02}
{Corbel}, S., \& {Fender}, R.~P.,  2002, ApJ, 573, L35

\bibitem[\protect\astroncite{{Corbel} et~al.}{2004}]{corbel04}
{Corbel}, S., {Fender}, R.~P., {Tomsick}, J.~A., {Tzioumis}, A.~K., \&
  {Tingay}, S.,  2004, ApJ, 617, 1272

\bibitem[\protect\astroncite{{Corbel} et~al.}{2000}]{corbel00}
{Corbel}, S., {Fender}, R.~P., {Tzioumis}, A.~K., {Nowak}, M., {McIntyre}, V.,
  {Durouchoux}, P., \& {Sood}, R.,  2000, A\&A, 359, 251

\bibitem[\protect\astroncite{{Cui}, {Chen} \& {Zhang}}{2000}]{cui00a}
{Cui}, W., {Chen}, W., \& {Zhang}, S.~N.,  2000, ApJ, 529, 952

\bibitem[\protect\astroncite{{Dauser} et~al.}{2013}]{dauser13}
{Dauser}, T., {Garcia}, J., {Wilms}, J., {B{\"o}ck}, M., {Brenneman}, L.~W.,
  {Falanga}, M., {Fukumura}, K., \& {Reynolds}, C.~S.,  2013, MNRAS, 430, 1694

\bibitem[\protect\astroncite{{Debnath}, {Chakrabarti} \&
  {Nandi}}{2013}]{debnath13}
{Debnath}, D., {Chakrabarti}, S.~K., \& {Nandi}, A.,  2013, Advances in Space
  Research, 52, 2143

\bibitem[\protect\astroncite{{del Santo} et~al.}{2005}]{delsanto05}
{del Santo}, M., et~al., 2005, A\&A, 433, 613

\bibitem[\protect\astroncite{{D{\'{\i}}az}~Trigo et~al.}{2013}]{diaztrigo13}
{D{\'{\i}}az}~Trigo, M., {Miller-Jones}, J.~C.~A., {Migliari}, S., {Broderick},
  J.~W., \& {Tzioumis}, T.,  2013, Nature, 504, 260

\bibitem[\protect\astroncite{{Dieters} et~al.}{2000}]{dieters00}
{Dieters}, S.~W., et~al., 2000, ApJ, 538, 307

\bibitem[\protect\astroncite{{Dunn} et~al.}{2010}]{dunn10}
{Dunn}, R.~J.~H., {Fender}, R.~P., {K{\"o}rding}, E.~G., {Belloni}, T., \&
  {Cabanac}, C.,  2010, MNRAS, 403, 61

\bibitem[\protect\astroncite{{Esin}, {McClintock} \& {Narayan}}{1997}]{emn97}
{Esin}, A.~A., {McClintock}, J.~E., \& {Narayan}, R.,  1997, ApJ, 489, 865

\bibitem[\protect\astroncite{{Fabian} et~al.}{1989}]{fabian89}
{Fabian}, A.~C., {Rees}, M.~J., {Stella}, L., \& {White}, N.~E.,  1989, MNRAS,
  238, 729

\bibitem[\protect\astroncite{{Fender} et~al.}{1999}]{fender99}
{Fender}, R., et~al., 1999, ApJ, 519, L165

\bibitem[\protect\astroncite{{Fender}}{2001}]{fender01}
{Fender}, R.~P.,  2001, MNRAS, 322, 31

\bibitem[\protect\astroncite{{Fender}, {Belloni} \& {Gallo}}{2004}]{fbg04}
{Fender}, R.~P., {Belloni}, T.~M., \& {Gallo}, E.,  2004, MNRAS, 355, 1105

\bibitem[\protect\astroncite{{Fender}, {Gallo} \& {Jonker}}{2003}]{fgj03}
{Fender}, R.~P., {Gallo}, E., \& {Jonker}, P.~G.,  2003, MNRAS, 343, L99

\bibitem[\protect\astroncite{{Fender}, {Homan} \& {Belloni}}{2009}]{fhb09}
{Fender}, R.~P., {Homan}, J., \& {Belloni}, T.~M.,  2009, MNRAS, 396, 1370

\bibitem[\protect\astroncite{{Garcia} et~al.}{2001}]{garcia01}
{Garcia}, M.~R., {McClintock}, J.~E., {Narayan}, R., {Callanan}, P., {Barret},
  D., \& {Murray}, S.~S.,  2001, ApJ, 553, L47

\bibitem[\protect\astroncite{{Garmire} et~al.}{2003}]{garmire03}
{Garmire}, G.~P., {Bautz}, M.~W., {Ford}, P.~G., {Nousek}, J.~A., \& {Ricker},
  G.~R.,  2003,
\newblock in X-Ray and Gamma-Ray Telescopes and Instruments for Astronomy.
  Edited by Joachim E. Truemper, Harvey D. Tananbaum. Proceedings of the SPIE,
  4851, 28

\bibitem[\protect\astroncite{{Hannikainen} et~al.}{2002}]{hannikainen02}
{Hannikainen}, D., {Sault}, B., {Kuulkers}, E., {Wu}, K., {Jones}, P., \&
  {Hunstead}, R.,  2002, The Astronomer's Telegram, 108

\bibitem[\protect\astroncite{{Hjellming} et~al.}{1999}]{hjellming99}
{Hjellming}, R.~M., et~al., 1999, ApJ, 514, 383

\bibitem[\protect\astroncite{{Homan} \& {Belloni}}{2005}]{hb05}
{Homan}, J., \& {Belloni}, T.,  2005, Ap\&SS, 300, 107

\bibitem[\protect\astroncite{{Kalemci} et~al.}{2013}]{kalemci13}
{Kalemci}, E., {Din{\c c}er}, T., {Tomsick}, J.~A., {Buxton}, M.~M., {Bailyn},
  C.~D., \& {Chun}, Y.~Y.,  2013, ApJ, 779, 95

\bibitem[\protect\astroncite{{Kalemci} et~al.}{2003}]{kalemci03a}
{Kalemci}, E., {Tomsick}, J.~A., {Rothschild}, R.~E., {Pottschmidt}, K.,
  {Corbel}, S., {Wijnands}, R., {Miller}, J.~M., \& {Kaaret}, P.,  2003, ApJ,
  586, 419

\bibitem[\protect\astroncite{{King} et~al.}{2014}]{king14}
{King}, A.~L., et~al., 2014, ApJ, 784, L2

\bibitem[\protect\astroncite{{Koyama} et~al.}{2007}]{koyama07}
{Koyama}, K., et~al., 2007, PASJ, 59, 23

\bibitem[\protect\astroncite{{Kubota} et~al.}{2007}]{kubota07}
{Kubota}, A., et~al., 2007, PASJ, 59, 185

\bibitem[\protect\astroncite{{Kuulkers} et~al.}{1997}]{kuulkers97}
{Kuulkers}, E., {Parmar}, A.~N., {Kitamoto}, S., {Cominsky}, L.~R., \& {Sood},
  R.~K.,  1997, MNRAS, 291, 81

\bibitem[\protect\astroncite{{Laor}}{1991}]{laor91}
{Laor}, A.,  1991, ApJ, 376, 90

\bibitem[\protect\astroncite{{Lightman} \& {White}}{1988}]{lw88}
{Lightman}, A.~P., \& {White}, T.~R.,  1988, ApJ, 335, 57

\bibitem[\protect\astroncite{{Liu}, {Meyer} \&
  {Meyer-Hofmeister}}{2005}]{liu05}
{Liu}, B.~F., {Meyer}, F., \& {Meyer-Hofmeister}, E.,  2005, A\&A, 442, 555

\bibitem[\protect\astroncite{{Maccarone}}{2003}]{maccarone03}
{Maccarone}, T.~J.,  2003, A\&A, 409, 697

\bibitem[\protect\astroncite{{McClintock} \& {Remillard}}{2006}]{mr06}
{McClintock}, J.~E., \& {Remillard}, R.~A.,  2006,
\newblock {Black hole binaries},
\newblock  Compact stellar X-ray sources.~Edited by Walter Lewin \& 
Michiel van der Klis: Cambridge University Press,  157--213

\bibitem[\protect\astroncite{{McClintock} et~al.}{2006}]{mcclintock06}
{McClintock}, J.~E., {Shafee}, R., {Narayan}, R., {Remillard}, R.~A., {Davis},
  S.~W., \& {Li}, L.-X.,  2006, ApJ, 652, 518

\bibitem[\protect\astroncite{{Menou} et~al.}{1999}]{menou99}
{Menou}, K., {Esin}, A.~A., {Narayan}, R., {Garcia}, M.~R., {Lasota}, J., \&
  {McClintock}, J.~E.,  1999, ApJ, 520, 276

\bibitem[\protect\astroncite{{Meyer-Hofmeister}, {Liu} \&
  {Meyer}}{2005}]{mlm05}
{Meyer-Hofmeister}, E., {Liu}, B.~F., \& {Meyer}, F.,  2005, A\&A, 432, 181

\bibitem[\protect\astroncite{{Miller}}{2007}]{miller07}
{Miller}, J.~M.,  2007, ARA\&A, 45, 441

\bibitem[\protect\astroncite{{Miller} et~al.}{2008}]{miller08}
{Miller}, J.~M., et~al., 2008, ApJ, 679, L113

\bibitem[\protect\astroncite{{Mitsuda} et~al.}{1984}]{mitsuda84}
{Mitsuda}, K., et~al., 1984, PASJ, 36, 741

\bibitem[\protect\astroncite{{Narayan}, {Garcia} \& {McClintock}}{1997}]{ngm97}
{Narayan}, R., {Garcia}, M.~R., \& {McClintock}, J.~E.,  1997, ApJ, 478, L79

\bibitem[\protect\astroncite{{Narayan} \& {McClintock}}{2008}]{nm08}
{Narayan}, R., \& {McClintock}, J.~E.,  2008, New Astronomy Review, 51, 733

\bibitem[\protect\astroncite{{Neilsen} et~al.}{2014}]{neilsen14}
{Neilsen}, J., {Coriat}, M., {Fender}, R., {Lee}, J.~C., {Ponti}, G.,
  {Tzioumis}, A.~K., {Edwards}, P.~G., \& {Broderick}, J.~W.,  2014, ApJ, 784,
  L5

\bibitem[\protect\astroncite{{Parmar}, {Angelini} \& {White}}{1995}]{paw95}
{Parmar}, A.~N., {Angelini}, L., \& {White}, N.~E.,  1995, ApJ, 452, L129

\bibitem[\protect\astroncite{{Parmar}, {Stella} \& {White}}{1986}]{psw86}
{Parmar}, A.~N., {Stella}, L., \& {White}, N.~E.,  1986, ApJ, 304, 664

\bibitem[\protect\astroncite{{Parmar} et~al.}{1997}]{parmar97}
{Parmar}, A.~N., {Williams}, O.~R., {Kuulkers}, E., {Angelini}, L., \& {White},
  N.~E.,  1997, A\&A, 319, 855

\bibitem[\protect\astroncite{{Petrucci} et~al.}{2008}]{petrucci08}
{Petrucci}, P.-O., {Ferreira}, J., {Henri}, G., \& {Pelletier}, G.,  2008,
  MNRAS, 385, L88

\bibitem[\protect\astroncite{{Plotkin}, {Gallo} \& {Jonker}}{2013}]{plotkin13}
{Plotkin}, R.~M., {Gallo}, E., \& {Jonker}, P.~G.,  2013, \apj, 773, 59

\bibitem[\protect\astroncite{{Reynolds} \& {Miller}}{2011}]{rm11}
{Reynolds}, M.~T., \& {Miller}, J.~M.,  2011, ApJ, 734, L17

\bibitem[\protect\astroncite{{Reynolds} et~al.}{2014}]{reynolds14}
{Reynolds}, M.~T., {Reis}, R.~C., {Miller}, J.~M., {Cackett}, E.~M., \&
  {Degenaar}, N.,  2014, MNRAS, 441, 3656

\bibitem[\protect\astroncite{{Ross} \& {Fabian}}{2005}]{rf05}
{Ross}, R.~R., \& {Fabian}, A.~C.,  2005, MNRAS, 358, 211

\bibitem[\protect\astroncite{{R{\'o}{\.z}a{\'n}ska} et~al.}{2014}]{rozanska14}
{R{\'o}{\.z}a{\'n}ska}, A., {Madej}, J., {Bagi{\'n}ska}, P., {Hryniewicz}, K.,
  \& {Handzlik}, B.,  2014, A\&A, 562, A81

\bibitem[\protect\astroncite{{Russell} et~al.}{2013}]{russell13}
{Russell}, D.~M., et~al., 2013, MNRAS, 429, 815

\bibitem[\protect\astroncite{{Russell} et~al.}{2011}]{russell11}
{Russell}, D.~M., {Miller-Jones}, J.~C.~A., {Maccarone}, T.~J., {Yang}, Y.~J.,
  {Fender}, R.~P., \& {Lewis}, F.,  2011, ApJ, 739, L19

\bibitem[\protect\astroncite{{Smith}, {Dawson} \& {Swank}}{2007}]{smith07}
{Smith}, D.~M., {Dawson}, D.~M., \& {Swank}, J.~H.,  2007, ApJ, 669, 1138

\bibitem[\protect\astroncite{{Smith} et~al.}{2001}]{smith01}
{Smith}, D.~M., {Heindl}, W.~A., {Markwardt}, C.~B., \& {Swank}, J.~H.,  2001,
  ApJ, 554, L41

\bibitem[\protect\astroncite{{Tomsick} et~al.}{2005}]{tomsick05c}
{Tomsick}, J.~A., {Corbel}, S., {Goldwurm}, A., \& {Kaaret}, P.,  2005, ApJ,
  630, 413

\bibitem[\protect\astroncite{{Tomsick} \& {Kaaret}}{2000}]{tk00}
{Tomsick}, J.~A., \& {Kaaret}, P.,  2000, ApJ, 537, 448

\bibitem[\protect\astroncite{{Tomsick} \& {Yamaoka}}{2010}]{ty10}
{Tomsick}, J.~A., \& {Yamaoka}, K.,  2010, The Astronomer's Telegram, 2794

\bibitem[\protect\astroncite{{Tomsick} et~al.}{2009}]{tomsick09c}
{Tomsick}, J.~A., {Yamaoka}, K., {Corbel}, S., {Kaaret}, P., {Kalemci}, E., \&
  {Migliari}, S.,  2009, ApJ, 707, L87

\bibitem[\protect\astroncite{{Verner} et~al.}{1996}]{vern96}
{Verner}, D.~A., {Ferland}, G.~J., {Korista}, K.~T., \& {Yakovlev}, D.~G.,
  1996, ApJ, 465, 487

\bibitem[\protect\astroncite{{Wilms}, {Allen} \& {McCray}}{2000}]{wam00}
{Wilms}, J., {Allen}, A., \& {McCray}, R.,  2000, ApJ, 542, 914

\end{thebibliography}

\clearpage

\begin{table}
\caption{Observations of 4U 1630--47\label{tab:obs}}
\begin{minipage}{\linewidth}
\begin{center}
\begin{tabular}{cccccc} \hline \hline
Satellite & Observation & ObsID & Start Time & End Time & Exposure Time\\
          & Number      &       & (UT)       & (UT)     & (s)\\ \hline\hline
{\em Swift}   & 1  & 00031224006 & 2010 Jul 15, 12.91 h & 2010 Jun 15, 16.59 h & 4,617\\
     ''       & 2  & 00031224007 & 2010 Aug 17, 1.90 h  & 2010 Aug 17, 2.09 h  & 690\\
     ''       & 3  & 00031224008 & 2010 Aug 19, 16.21 h & 2010 Aug 19, 16.54 h & 1,149\\
     ''       & 4  & 00031224009 & 2010 Aug 22, 0.46 h  & 2010 Aug 22, 3.96 h & 1,348\\
{\em Suzaku}  & 5  & 405051010   & 2010 Aug 24, 21.88 h & 2010 Aug 27, 14.72 h & 99,940\\
{\em Chandra} & -- & 12530       & 2011 Jun 16, 12.53 h & 2011 Jun 16, 18.15 h & 19,260\\
     ''       & -- & 12533       & 2011 Jun 17, 5.55 h  & 2011 Jun 17, 11.26 h & 19,503\\ \hline
\end{tabular}
\end{center}
\end{minipage}
\end{table}

\begin{table}
\caption{Spectral Parameters from Power-law Fits\label{tab:parameters}}
\begin{minipage}{\linewidth}
\begin{center}
\begin{tabular}{ccccc} \hline \hline
Observation  & $N_{\rm H}$\footnote{To calculate the column density, we use \cite{wam00} abundances and \cite{vern96} cross-sections.}             & $\Gamma$  & Flux\footnote{Unabsorbed 2--10\,keV flux in units of $10^{-12}$\,erg\,cm$^{-2}$\,s$^{-1}$.} & $\chi^{2}_{\nu}$/dof\\
Number       & ($10^{22}$\,cm$^{-2}$)  &                  &         &  \\\hline\hline
1 & $12\pm 3$\footnote{Throughout this table, we quote 90\% confidence errors.}    & $4.3\pm 1.0$      & $9^{+6}_{-3}$     & 0.57/12\\
2 & $7\pm 2$     & $1.6\pm 0.4$      & $112^{+13}_{-11}$ & 1.26/34\\
3 & $11\pm 3$    & $1.7\pm 0.5$      & $74^{+13}_{-9}$   & 1.07/34\\
4 & $9^{+5}_{-4}$ & $1.4\pm 0.7$      & $25^{+6}_{-4}$   & 1.18/34\\
5\footnote{For {\em Suzaku}, we left the overall normalization of XIS1 relative to XIS0/3 as a free parameter, and we obtain a value of $1.09\pm 0.03$.} & $5.8\pm 0.3$ & $1.58\pm 0.05$    & $5.94\pm 0.11$  & 0.99/185\\ \hline\hline
\end{tabular}
\end{center}
\end{minipage}
\end{table}

\begin{table}
\caption{Comparison between {\em Swift}/XRT and {\em RXTE}/PCA\label{tab:confusion}}
\begin{minipage}{\linewidth}
\begin{center}
\begin{tabular}{cccccc} \hline \hline
{\em Swift} & ObsID of           & {\em RXTE} & Duration     & Predicted  & Actual\\
Observation & Closest {\em RXTE} & Start Time & Between      & PCA Rate\footnote{The count rate prediction for one PCU based on the absorbed power-law model measured by {\em Swift}. The values in parentheses are the predicted rates, including the exponential decay of 4U~1630--47. }   & PCA Rate\\
Number      & Observation        & (UT)       & Start Times  & (3-25 keV) & (3-25 keV)\\ \hline\hline
1  &  95702-03-14-03 & 2010 July 15, 20.1 h & 7.2 h  & 0.23 c/s  & 10.3 c/s\\
2  &  95702-03-21-01 & 2010 Aug 17, 3.1 h   & 1.2 h  & 11.6 (11.4) c/s  & 23.5 c/s\\
3  &  95702-03-22-00 & 2010 Aug 20, 18.0 h  & 25.8 h & 6.8 (5.0) c/s   & 17.5 c/s\\
4  &  95702-03-22-01 & 2010 Aug 23, 3.3 h   & 26.8 h & 2.8 (2.0) c/s  & 17.6 c/s\\ \hline
\end{tabular}
\end{center}
\end{minipage}
\end{table}

\end{document}